\documentclass[aps,showpacs,preprint,superscriptaddress]{revtex4}
\usepackage{graphicx}
\usepackage{subfigure}
\usepackage{float}
\usepackage{amsmath}
\usepackage{amsfonts}
\usepackage{bm}
\usepackage{bbm}
\usepackage{txfonts}
\usepackage{array}
\usepackage{amssymb}

\begin{document}

\title{The high density $\gamma$-ray emission and dense positron production via multi-lasers driven circular target}
\author{Ya-Juan Hou}
\affiliation{Key Laboratory of Beam Technology of the Ministry of Education, and College of Nuclear Science and Technology, Beijing Normal University, Beijing 100875, China}
\author{Bai-Song Xie \footnote{bsxie@bnu.edu.cn}}
\affiliation{Key Laboratory of Beam Technology of the Ministry of Education, and College of Nuclear Science and Technology, Beijing Normal University, Beijing 100875, China}
\affiliation{Beijing Radiation Center, Beijing 100875, China}
\author{Chong Lv}
\affiliation{Key Laboratory of Beam Technology of the Ministry of Education, and College of Nuclear Science and Technology, Beijing Normal University, Beijing 100875, China}
\author{Feng Wan}
\affiliation{Key Laboratory of Beam Technology of the Ministry of Education, and College of Nuclear Science and Technology, Beijing Normal University, Beijing 100875, China}
\author{Li Wang}
\affiliation{Key Laboratory of Beam Technology of the Ministry of Education, and College of Nuclear Science and Technology, Beijing Normal University, Beijing 100875, China}
\author{Nureli Yasen}
\affiliation{Key Laboratory of Beam Technology of the Ministry of Education, and College of Nuclear Science and Technology, Beijing Normal University, Beijing 100875, China}
\author{Hai-Bo Sang}
\affiliation{Key Laboratory of Beam Technology of the Ministry of Education, and College of Nuclear Science and Technology, Beijing Normal University, Beijing 100875, China}
\author{Guo-Xing Xia}
\affiliation{School of Physics and Astronomy, University of Manchester, Manchester M13 9PL, United Kingdom}
\affiliation{The Cockcroft Institute, Warrington, WA4 4AD, United Kingdom}
\date{\today}

\begin{abstract}
A diamond-like carbon circular target is proposed to improve the $\gamma$-ray emission and pair production with lasers intensity of $8\times 10^{22} ~\mathrm{W/cm^2}$ by using two-dimensional particle-in-cell simulations with quantum electrodynamics. It is found that the circular target can significantly enhance the density of $\gamma$-photons than plane target when two colliding circularly polarized lasers irradiate the target. By multi-lasers irradiate the circular target, the optical trap of lasers can prevent the high energy electrons accelerated by laser radiation pressure from escaping. Hence, high density as $5164 n_c$ $\gamma$-photons is obtained through nonlinear Compton back-scattering. Meanwhile, $2.7 \times 10^{11}$ positrons with average energy of $230 ~\mathrm{MeV}$ is achieved via multi-photon Breit-Wheeler process. Such ultrabright $\gamma$-ray source and dense positrons source can be useful to many applications. The optimal target radius and laser mismatching deviation parameters are also discussed in detail.

\end{abstract}
\maketitle

\section{\label{Introduction}Introduction}

With the rapid development of laser technologies, laser intensity of $10^{22} ~\mathrm{W/cm^2}$ has been demonstrated \cite{Yanovsky}. Extreme laser intensity like $10^{23} ~\mathrm{W/cm^2}$ is available in the next few years, which means the electron dynamics approaching nonlinear quantum electrodynamics (QED) regime \cite{Dipiazza,Mourou}. Such laser intensity will allow studying bright $\gamma$-ray emission, $e^+e^-$ pair production, QED-cascade and particles acceleration in laboratories \cite{Remington,Rufni}. Intense $\gamma$-ray sources are useful for simulating the celestial process and extreme environments \cite{Aharonian}. In the past decades, many researches focused on the $\gamma$-ray emission and pair production \cite{Avetissian,Shen,Chen,Liang,Avetissian2,Shkolnikova,Berezhiani}. At extremely high laser intensity, nonlinear Compton scattering is an important way for $\gamma$-ray emission through colliding relativistic electrons with intense laser pulse \cite{Di Piazza,Ta Phuoc,Sarri,Sakai}. This high energy $\gamma$-photons colliding with lasers enables the laser energy to convert into $e^+e^-$ pairs via multi-photon Breit-Wheeler (BW) process \cite{Breit,Nikishov}.

Several schemes are proposed to generate bright $\gamma$-ray and pair production via nonlinear Compton scattering and BW process. Among them, one way is to enhance the laser intensity by selecting appropriate polarized lasers \cite{Gelfer,Yuan,Marija} or/and focusing and redistributing the lasers energy\cite{Bulanov,Gonoskov,Esirkepov,Kirk,Vranic}. Another way is to change the plasma target configuration, such as one or multiple laser interaction with near-critical-density plasma \cite{JIN-JIN LIU,Huang,Zhu,Brady}, solid Al target \cite{Ridgers1,Luo,Chang} or gas plasma \cite{Liu,Lobet}. Among them, laser wakefield acceleration \cite{Liu J X} and laser ponderomotive acceleration \cite{HAN-ZHEN LI,Hu} are generally used to enhance electron acceleration and constraint. Recently, the radiation pressure acceleration (RPA) of ultra-thin foils is also applied to $\gamma$-ray emission and dense $e^+e^-$ pairs production \cite{Lihanzhen}, as it is capable of obtaining high energy electrons and quasi-monoenergetic ion beams \cite{Henig,Macchi,Lv,Xueren}. However, the laser intensity $5 \times 10^{23} ~\mathrm{W/cm^2}$ is too high to obtain experimentally and on the other hand, the plane target cannot prevent the electrons from transverse escaping.

In this paper, a diamond-like carbon (DLC) circular target is presented as an alternative to prevent the electrons from escaping transversely. It is obvious that when the circular target is used, laser energy conversion efficiency to $\gamma$-photons is enhanced and the $\gamma$-photons number density is about twice higher than that of the plane target. Besides, the circular target allows an interaction with multi-lasers at the same time, the optical trap generated in situ can reduce the electrons escaping more efficiently. Eventually, an ultrabright $\gamma$-ray emission with a high density of $5164 n_c$ is obtained at $14 T_0$ (where $T_0$ is the laser period) under the laser intensity $8 \times 10^{22} ~\mathrm{W/cm^{2}}$ through the nonlinear Compton back-scattering (NCBS) process. Further these $7.5 \times 10^{14}$ photons with the average energy of $16 ~\mathrm{MeV}$ colliding with lasers can produce dense positrons with more than $20 n_c$ density via multi-photon BW process. The total positron yield can be as high as $2.7 \times 10^{11}$, with average energy is about $230 ~\mathrm{MeV}$.

The paper is organized as follows. Section 2 outlines the basic target configurations and simulation parameters. The $\gamma$-ray emission by two circularly polarized (CP) laser-driven target is also discussed in detail. Section 3 examines the ultrabright $\gamma$-ray emission and $e^+e^-$ pair production through RPA by four CP lasers irradiating a circular target. Among them, the optimal target radius and the deviation of lasers mismatching are also taken into account. Lastly, a brief summary is given in Sec. 4.

\section{Ultrabright $\gamma$-ray emission by two lasers irradiating a circular target}

The 2D3V simulation results of ultrabright $\gamma$-ray emission by two laser-driven DLC target is performed via QED-PIC code EPOCH \cite{Ridgers,Duclous}.

\begin{figure*}[htbp]\suppressfloats
\includegraphics[width=17cm, height=10cm]{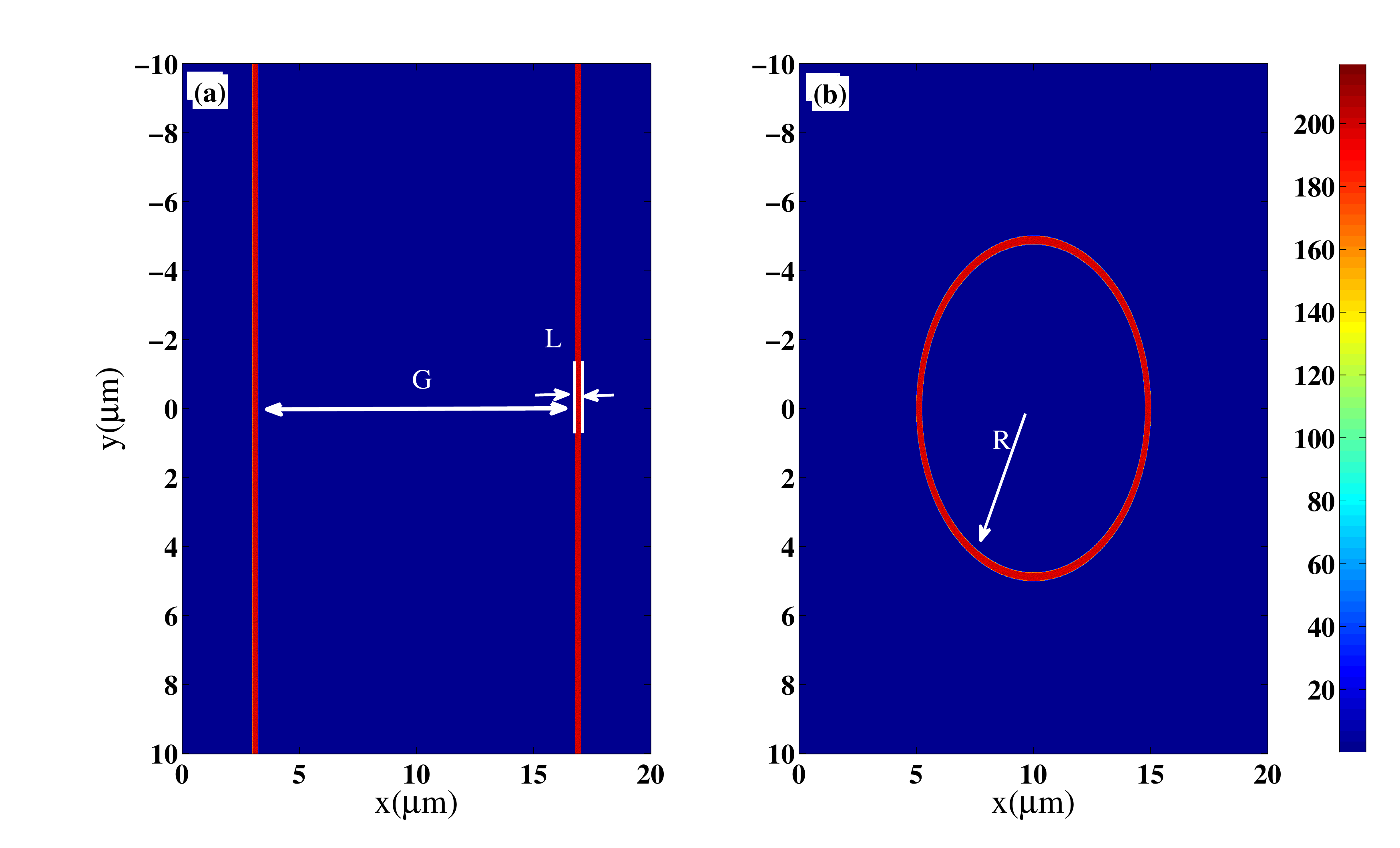}
\caption{\label{target} (Color online) Simulation box and initial target structure. Initial plasma density of plane target (a) and circular (b). The density is normalized by the critical density $n_c$.}
\label{fig:1}
\end{figure*}

The DLC foils are ideal materials for self-supporting targets in experiments due to it's high tensile strength, hardness and heat resistance \cite{Liechtenstein}. In our scheme, a circular DLC target as shown in Fig.\ref{target}(b) is used instead of the plane DLC target as shown in Fig.\ref{target}(a) \cite{Lihanzhen} to get brighter $\gamma$-ray and denser $e^+e^-$ pairs through RPA. The DLC target is a plasma consisting of electrons, protons and full ionized carbon ions with charge state $Z_i=6$ and mass $m_i=12\times 1836 m_e$, where $m_e$ is the electron mass. The density of target is $n_e=200n_c$, mixed with $20 \%$ protons in number density, where $n_c=m_e\omega_0^2/4\pi e^2$ ($\omega_0$ is the frequency and $-e$ is the charge) is the critical density of plasma. As Fig.\ref{target} shows, the simulation box size is $20\lambda \times 20\lambda$ with $2000$ $\times$ $1400$ grid cells. Two identical CP laser pulses are incident from the center of left and right boundary of the box simultaneously. Each laser has a peak intensity of $8\times 10^{22} ~\mathrm{W/cm^2}$ and rises in about 1 $T_0$ and then keeps the maximum amplitude for $9$ $T_0$, where $T_0=\lambda/c$ is the laser period, $\lambda=1 ~\mathrm{\mu m}$ is the wavelength of laser and $c$ is the speed of light. The laser is Gaussian profile in $y$ direction with a spot size of ${4 ~\mathrm{\mu m}}$ [full width at half maximum (FWHM)]. When the laser intensity is $8\times 10^{22} ~\mathrm{W/cm^2}$, the optimal thickness and foil gap of plane target for $\gamma$-ray emission and pair production have been studied in detail \cite{Lihanzhen}. So, both targets in our scheme have a thickness of $L=0.25 ~\mathrm{\mu m}$ and the coordinates of the target centre is $(x, y)=(10 ~\mathrm{\mu m}, 0)$. The foil gap is $G=13.5 ~\mathrm{\mu m}$ for the target and the radius is $R=5 ~\mathrm{\mu m}$ for circular target. Note that, only the $\gamma$-photons whose energy is larger than $1 ~\mathrm{MeV}$ are counted in the following simulations.

\begin{figure*}[htbp]\suppressfloats
\includegraphics[width=17cm, height=10cm]{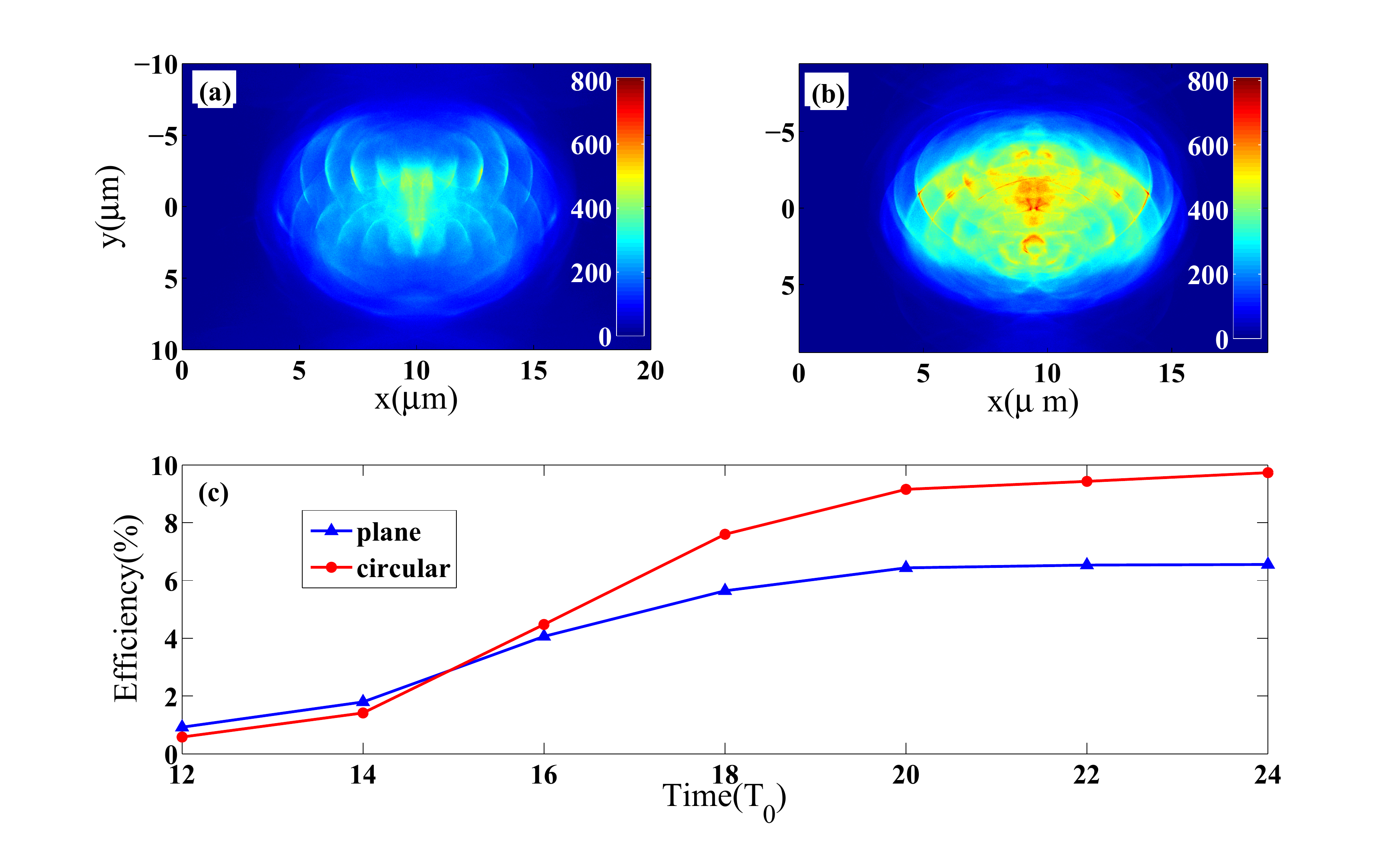}
\caption{\label{twolaser} (Color online) Distributions of photon density of plane target (a) and circular target (b) at $20 T_0$. Laser energy conversion efficiency to $\gamma$-photons (c). The density is normalized by the critical density $n_c$.}
\label{fig:1}
\end{figure*}

In the initial stage, the electrons, carbon ions and protons are separated and form a big charge separation field resulting in an inefficient acceleration of electrons due to the heavier carbon ions by using of DLC target. As time goes on, most of the laser waves penetrates through the target and begins to collide with the relativistic electrons accelerated by opposite laser from the other side through RPA. At this point, high energy $\gamma$-photons is generated through NCBS. When the high energy $\gamma$-photons collide with the lasers, the $e^+e^-$ pairs is produced via multi-photon BW process. For the plane target, the laser intensity along the axis increases since the laser is further focused in the inner surfaces as the target undergoes significant deformation. So, a large number of electrons escape from the foil, which result in a low density of $\gamma$-ray, as shown in Fig.\ref{twolaser}(a). The circular target we proposed can enhance the $\gamma$-photons density to $800 n_c$ which is about $2$ times the $\gamma$-photons density of plane target. There are two reasons for this enhancement. On the one hand, the circular target structure can slow down the laser pulse focusing and laser intensity increasing which will reduce the electrons escaping. On the other hand, the lasers pull the electrons out of the circular target continually and replenish the electrons source when the foil deforms and $\gamma$-ray emits. Besides, the high density $\gamma$-ray, shown in Fig.\ref{twolaser}(b), can sustain about $5 T_0$ which may become a stable $\gamma$-ray source in the future laboratory.

The laser energy conversion efficiency to $\gamma$-photons for plane target (the blue triangle curve) and circular target (the red circular curve) is plotted in Fig.\ref{twolaser}(c). The energy conversion efficiency of laser-to-photons for the plane target is about $6 \%$, which is comparable with the 3D simulation result of Ref. \cite{Lihanzhen}. It is evident that the circular target can significantly enhance the energy conversion efficiency of laser-to-photons to about $9 \%$ as time goes on. However, compared with the plane target, the lower laser intensity caused by circular structure also reduces the cutoff energy of electrons and $\gamma$-photons at the same time.

While the lower $\gamma$-photons energy may reduce the possibility of $e^+e^-$ pairs production to some extent, the circular target irradiated by multiple lasers has still an obvious advantage that can be seen in the following study. It not only affords a stable and high density $\gamma$-ray source but also provides a chance to get higher density $\gamma$-photons and more $e^+e^-$ pairs.

\section{$\gamma$-ray emission and $e^+e^-$ pairs production by multi-lasers driven circular target}

In order to demonstrate the enhancement of ultrabright $\gamma$-ray emission and dense $e^+e^-$ pairs production by multi-lasers driven DLC circular target, we perform the 2D3V simulation using QED-PIC code EPOCH. The simulation parameters are the same as presented in section $\expandafter{2}$ except that two additional CP lasers are incident from the center of up and down boundary of the simulation box and these two lasers are Gaussian profile in $x$ direction.

\begin{figure*}[htbp]\suppressfloats
\includegraphics[width=17cm, height=10cm]{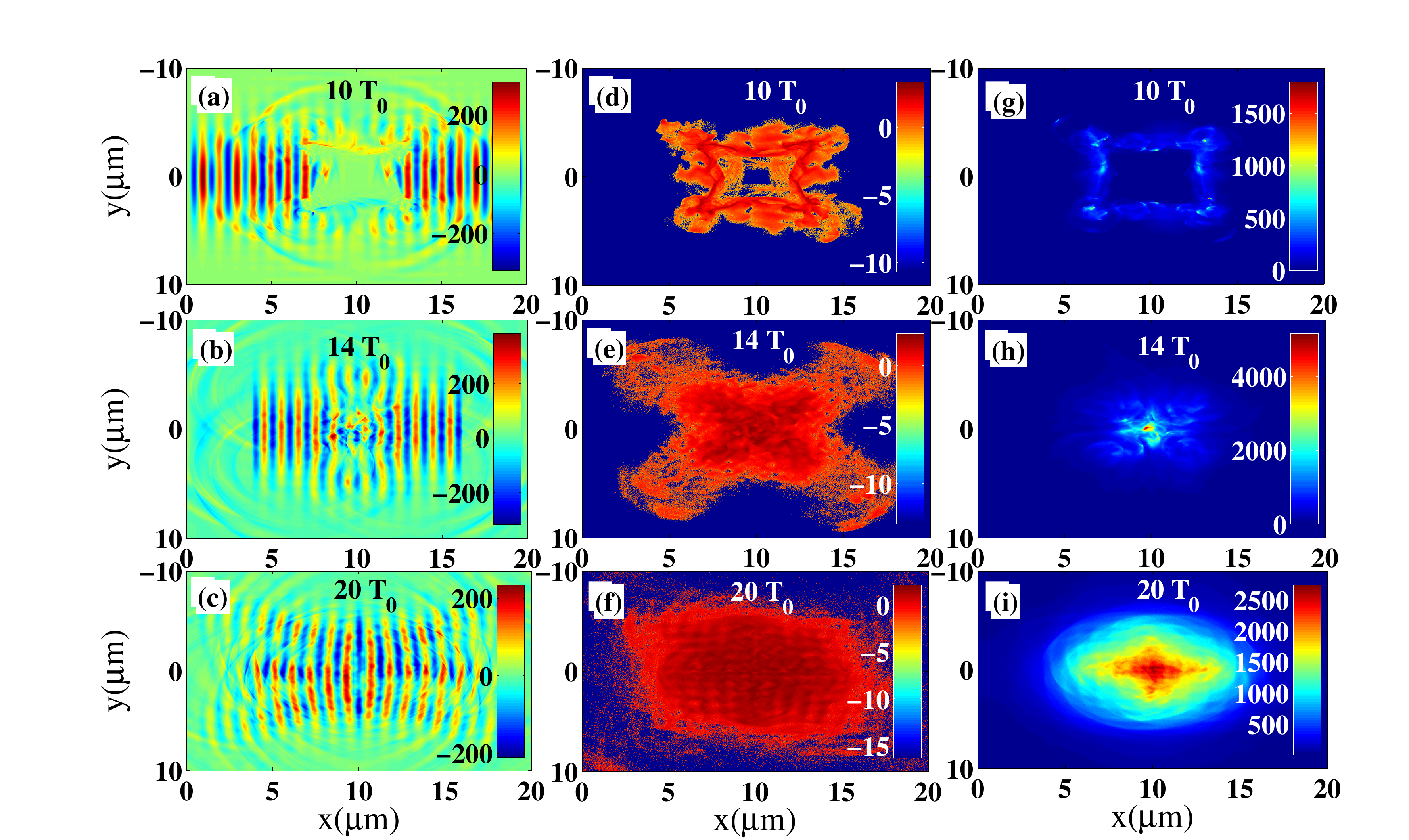}
\caption{\label{electrondensity}(Color online) Distributions of transverse electric field $E_y$ (a-c), electron density (d-f) and photon density (g-i) at $10 T_0$, $14 T_0$ and $20 T_0$, respectively. Here the electric field is normalized by $E_0=m_e\omega_0 c/e$ and the density is normalized by the critical density $n_c$.}
\label{fig:3}
\end{figure*}

\subsection{$\gamma$-ray emission}

Figure \ref{electrondensity} presents the transverse electric field (a-c), electrons density (d-f) and photons density (g-i) distribution of circular target at different stage. The probability rate for $\gamma$-ray emission in the QED regime is determined by a quantum invariant $\chi_{e^-} = (1/a_s)\sqrt{(\varepsilon_{e^-} E + P_{e^-} \times B)^2 - (P_{e^-} \cdot E)^2}$, where $a_s = eE_s/m_ec\omega_0 = m_ec^2/\hslash\omega_0$ is the normalized QED critical field, $E_s = m_ec^3/(\hslash e) = 1.32 \times 10^{18} ~\mathrm{Vm^{-1}}$ is the Schwinger field \cite{Schwinger}, $\varepsilon_{e^-} = \gamma_{e^-}m_ec^2$ is the electron energy, $\gamma_{e^-}$ is the Lorentz factor, $P_{e^-}$ is the electron momentum, $E$ and $B$ are the electromagnetic fields. Through analyzing, we know that $\chi_{e^-} \simeq 0$ and almost no high energy $\gamma$-photons is produced if the electrons interact with co-propagating lasers. When the electrons collide with the counter-propagating lasers, the quantum invariant $\chi_{e^-}$ can be $\chi_{e^-} \simeq 2\gamma_{e^-}E/E_s$. Hence, the $\gamma$-ray emission is generated if $\chi_{e^-} \geq 1$, which rely on the electrons energy and electric field intensity of lasers.

At the first stage, the initial circular target is distorted to be the four cone structures by the four lasers. Some electrons are first pulled out from the inner wall of target and then rapidly accelerated to high speed by the laser pressure and form overdense relativistic electron layers, as shown in Fig.\ref{electrondensity}(d). A big charge separation field is formed meanwhile due to the heavy protons and heavier carbon ions of DLC target materials which in turn pull ions forward. The accelerated electrons interact with the reflected laser waves resulting in $\gamma$-ray emission by NCBS, as seen in Fig.\ref{electrondensity}(g).

As shown in Fig.\ref{electrondensity}(b), the lasers in cone top are further focused and the intensity is enhanced when the target is expanding. So, the central residual electrons of the target are pulled off, as shown in Fig.\ref{electrondensity}(e), and the relativistic transparency of the DLC target occurs now, which means the lasers will penetrate through the target and collide with the counter-propagating electrons at about $14 T_0$. Through the NCBS, the $\gamma$-ray emission is enhanced resulting in a ultrabright $\gamma$-ray with a high peak density of $5164 n_c$, see Fig.\ref{electrondensity}(h). At $14 T_0$, for two-lasers driven DLC circular target, the photon peak density is $1914 n_c$. This indicates that the peak density of photons is increased about $2.7$ times when other two lasers are injected on side. In addition, the high density $\gamma$-photons can sustain about $20~\mathrm{fs}$. One reason for these benefits is the escaped electrons with transverse velocity will also interact with side lasers to realize the $\gamma$-ray emission enhancement. Another more important reason is that the optical traps created by multiple lasers prevent the electrons escaping from the region of maximum laser intensity.

\begin{figure*}[htbp]\suppressfloats
\includegraphics[width=17cm, height=10cm]{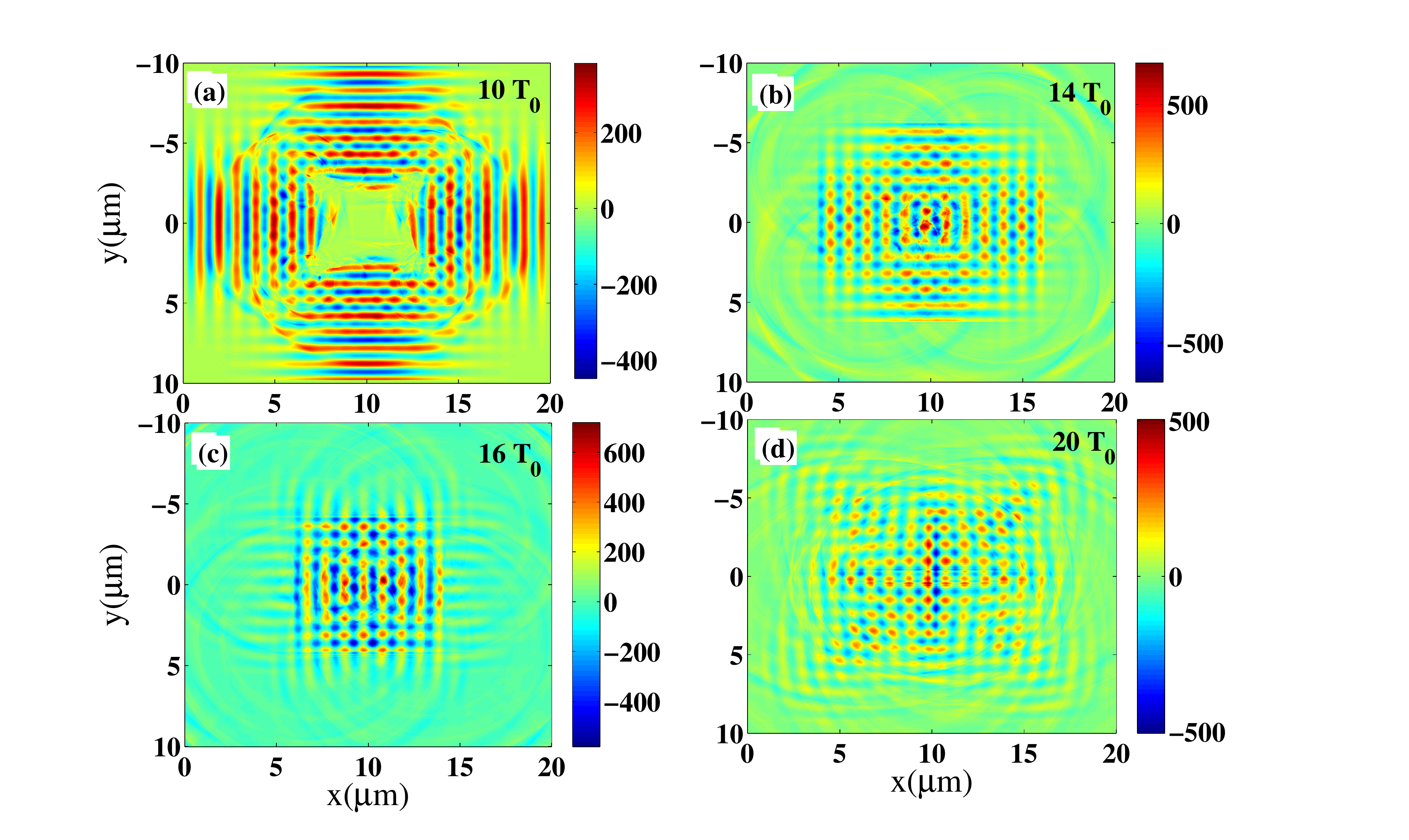}
\caption{\label{magnetic}(Color online) The distribution of transverse magnetic fields $B_z$ at $10 T_0$ (a), $14 T_0$ (b), $16 T_0$ (c) and $20 T_0$ (d). Here the transverse magnetic field is normalized by $m_e\omega c/e$.}
\label{fig:5}
\end{figure*}

Due to the limitation of our computer sources, here we only present a 2D optical trap with four lasers in Fig.\ref{magnetic}, which is a mimic of a real 3D optical trap formation by six laser beams. Some striking features are kept in our 2D simulations.  It was shown that the lattice-like magnetic field $B_z$ is formed and the intensity is enhanced as the lasers began to overlap at $14 T_0$. At $16 T_0$, the lasers are fully overlap and the magnetic field $B_z$ is enhanced from $300 ~\mathrm{MG}$ to about $700 ~\mathrm{MG}$, which is beneficial for trapping electrons and maximising the $\gamma$-ray emission. It is obvious that the different lattice-like optical trap structures at different stage correspond to different lattice structures of density distributions of electrons and $\gamma$-photons, which can be seen by comparing Fig.\ref{electrondensity} and Fig.\ref{magnetic}. This confirms our judgement more efficiently that the optical traps created by multiple lasers are the main reason for $\gamma$-ray emission enhancement.

\begin{figure*}[htbp]\suppressfloats
\includegraphics[width=17cm, height=10cm]{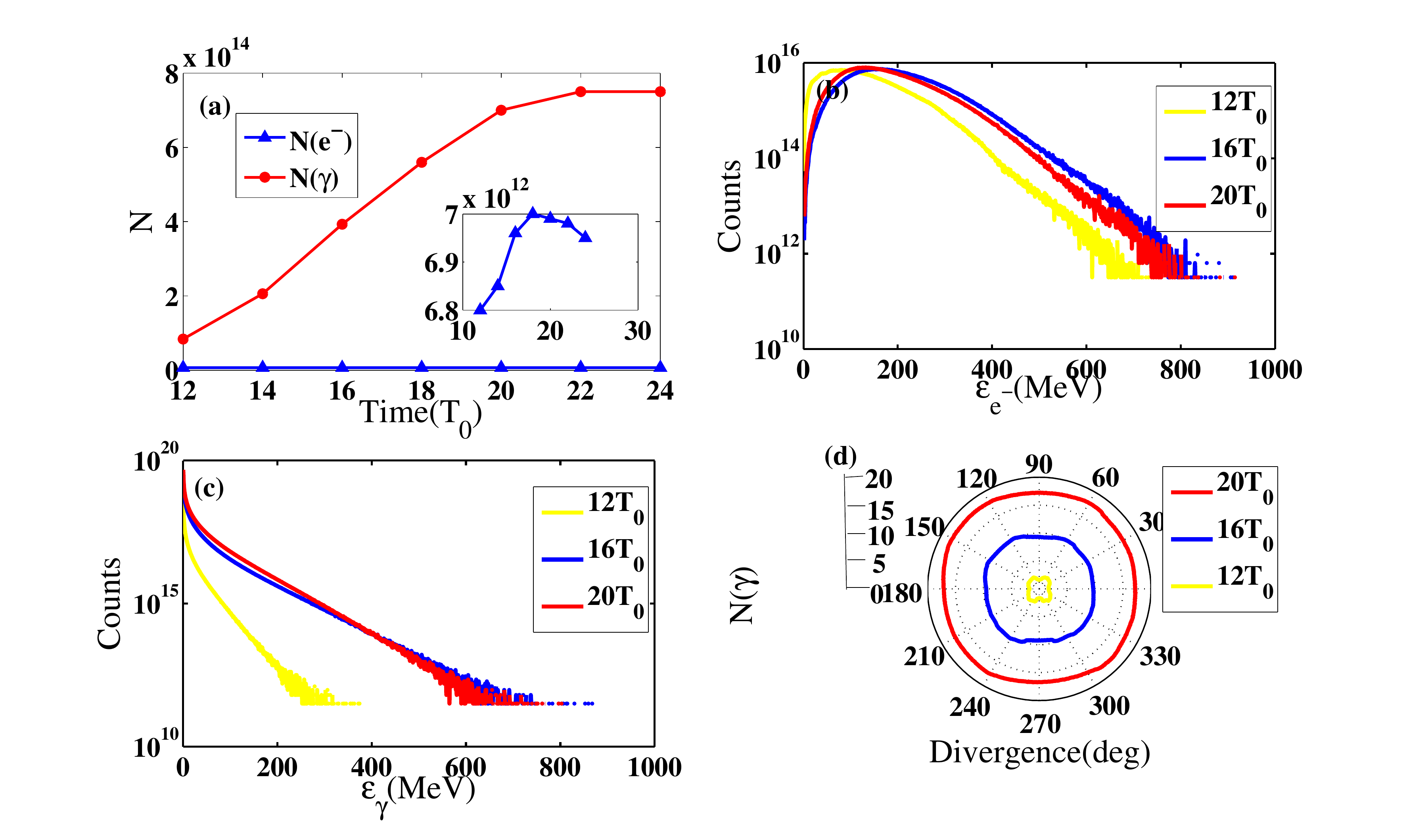}
\caption{\label{energy}(Color online) The numbers of electrons (blue triangle line) and photons (red roundness line) at different times (a), the energy spectrum of electrons (b), of $\gamma$-photons (c) and divergence distribution of photons (d) at $12 T_0$ (yellow line), $16 T_0$ (blue line) and $20 T_0$ (red line), respectively.}
\label{fig:4}
\end{figure*}

The lattice-like optical trap structure is diffused and the $B_z$ is reduced as well, as shown in Fig.\ref{magnetic}(d). In this last stage, after the lasers penetrating across the central intersection area, the overlapping area of lasers would be reduced gradually. The lasers push the electrons and $\gamma$-photons away from the center which results in a low number density of electrons and $\gamma$-photons, as shown in Fig.\ref{electrondensity}(i). The growth rate of electrons and $\gamma$-photons number at different times is presented in Fig.\ref{energy}(a). It can be seen from this that the total $\gamma$-photons number is $7.5 \times 10^{14}$, which is enhanced by over an order of magnitude compared to the $\gamma$-ray source in Ref. \cite{Lihanzhen}. Besides, the average energy of obtained $\gamma$-photons can be about $16 ~\mathrm{MeV}$, which can be seen in Fig.\ref{radius}(b).

Above all, $7.5 \times 10^{14}$ $\gamma$-photons with average energy $16 ~\mathrm{MeV}$ is obtained via multi-lasers driven DLC circular target. The maximum density of $\gamma$-ray can be $5164 n_c$ at $14 T_0$, which can be as an extremely dense and ultrabright $\gamma$-ray source for future application. This high quality photons will also have a significant benefit for pair production in BW process.

\subsection{Dense $e^+e^-$ pairs production}

In the QED region, multi-photon BW process is an very important mechanism for pair production through photon-photon  annihilation ($\gamma + n\hslash\omega_l \rightarrow e^- + e^+$). The probability for pair production via multi-photon BW process is determined by another quantum parameter $\chi_\gamma = (1/a_s)\sqrt{(\varepsilon_\gamma E + P_\gamma \times B)^2 - (P_\gamma \cdot E)^2} \simeq (2\hslash\omega_\gamma/m_ec^2)E/E_s$, here, $\varepsilon_\gamma = \hslash\omega_\gamma$, $P_\gamma = \hslash\omega_\gamma / c$ ($\omega_\gamma$ the photon frequency). So, the pair production depend on the photons energy $\hslash\omega_\gamma$ and electric field $E$ in interaction zone.

\begin{figure*}[htbp]\suppressfloats
\includegraphics[width=17cm, height=10cm]{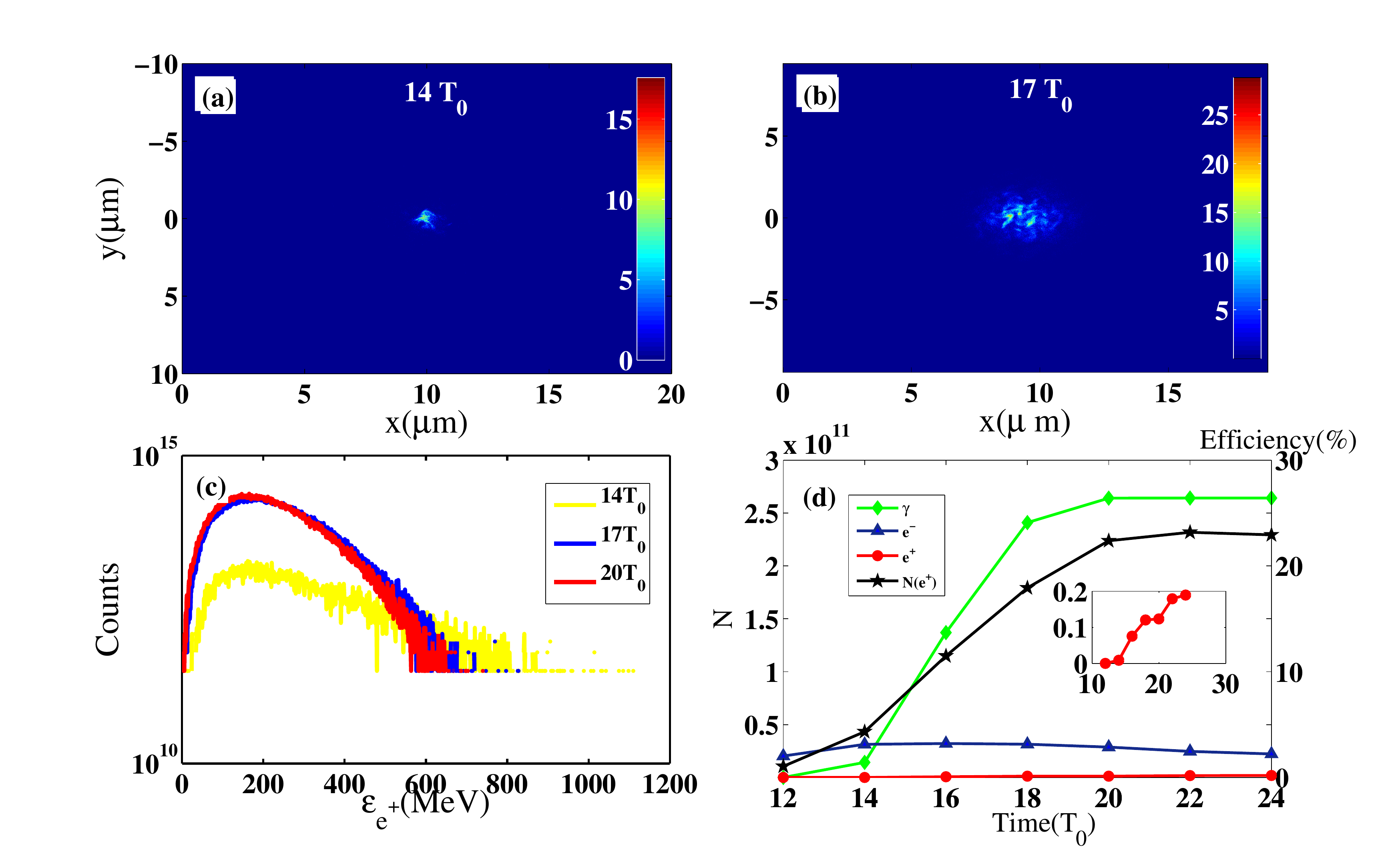}
\caption{\label{positrons}(Color online) The density distributions [(a) and (b)] and energy spectrum (c) of positrons. The laser energy conversion efficiency to electrons, to $\gamma$-photons
and to positrons as well as the number of positrons at different times (d).}
\label{fig:6}
\end{figure*}

Figure \ref{energy}(c) illustrates the energy spectrum of $\gamma$-photons at different times. Here, due to Doppler red shift, the reflected laser is weakened so that the maximal value of $\gamma$-photons is only $380 ~\mathrm{MeV}$ at the first stage, such as $12 T_0$, which is not enough to produce $e^+e^-$ pairs. At the second stage, the maximal cutoff energy of $\gamma$-photons can be $850 ~\mathrm{MeV}$ by NCBS while it will decrease since the high energy $\gamma$-photons are continually applied to BW process. The spectrum has a wide distribution and the average energy $\bar{\varepsilon_\gamma}$ can be $16 ~\mathrm{MeV}$ at $17 T_0$. Here the high photon energy can greatly enhance the possibility of pair production. As an example, Fig.\ref{positrons}(a) and \ref{positrons}(b) present the positrons density distribution at $14 T_0$ and $17 T_0$, respectively. At $14 T_0$, when high energy $\gamma$-photons and lasers collide, the positron yield starts being considerable. In second stage, the positron density remains at about $20 n_c$ and the maximum value can be $29 n_c$ at $17 T_0$, as Fig.\ref{positrons}(b) shows. These long-lasting and high-bunching positrons show a good prospect for potential applications in future.

The energy spectrum of positrons at different times has also be plotted in Fig.\ref{positrons}(c). The maximum energy of positrons obtained can be as high as $~\mathrm{GeV}$ at $14 T_0$. However, this higher energy positrons also oscillate in laser field and emit $\gamma$-ray resulting in a decrease of cutoff energy as time goes on. Beyond that, there is a monoenergetic peak at $200 ~\mathrm{MeV}$, which means monoenergetic positrons can be achieved through our scheme. The significant increase of positron number through this process is plotted in Fig.\ref{positrons}(d), which shows that the final number of positrons is $2.7 \times 10^{11}$. Besides the mean positron energy can be $230 ~\mathrm{MeV}$ at $17 T_0$. Note that the third dimensional size is assumed as $4 ~\mathrm{\mu m}$ according to the laser spot size and the positron distribution in order to a reasonable mimic of full 3D reality as much as possible.

To make the entire process more intuitive, we also calculate the time-dependent of laser efficiency to electrons, to $\gamma$-photons and to positrons. The general finding is that the laser energy conversion efficiency to $\gamma$-ray and positrons have a rapid growth in the second stage. As time goes on, the total laser energy conversion efficiency to $\gamma$-photons and positrons are about $27 \%$ and $0.2 \%$, respectively, which is a really high exploitation of laser energy.

\subsection{The effect of the target radius on $\gamma$-ray emission and $e^+e^-$ pairs production}

In our previous simulations, the circular target radius is chosen as $5 ~\mathrm{\mu m}$. Actually the radius of circular target plays an important role in $\gamma$-ray emission and pair production in real application.

\begin{figure*}[htbp]\suppressfloats
\includegraphics[width=17cm, height=10cm]{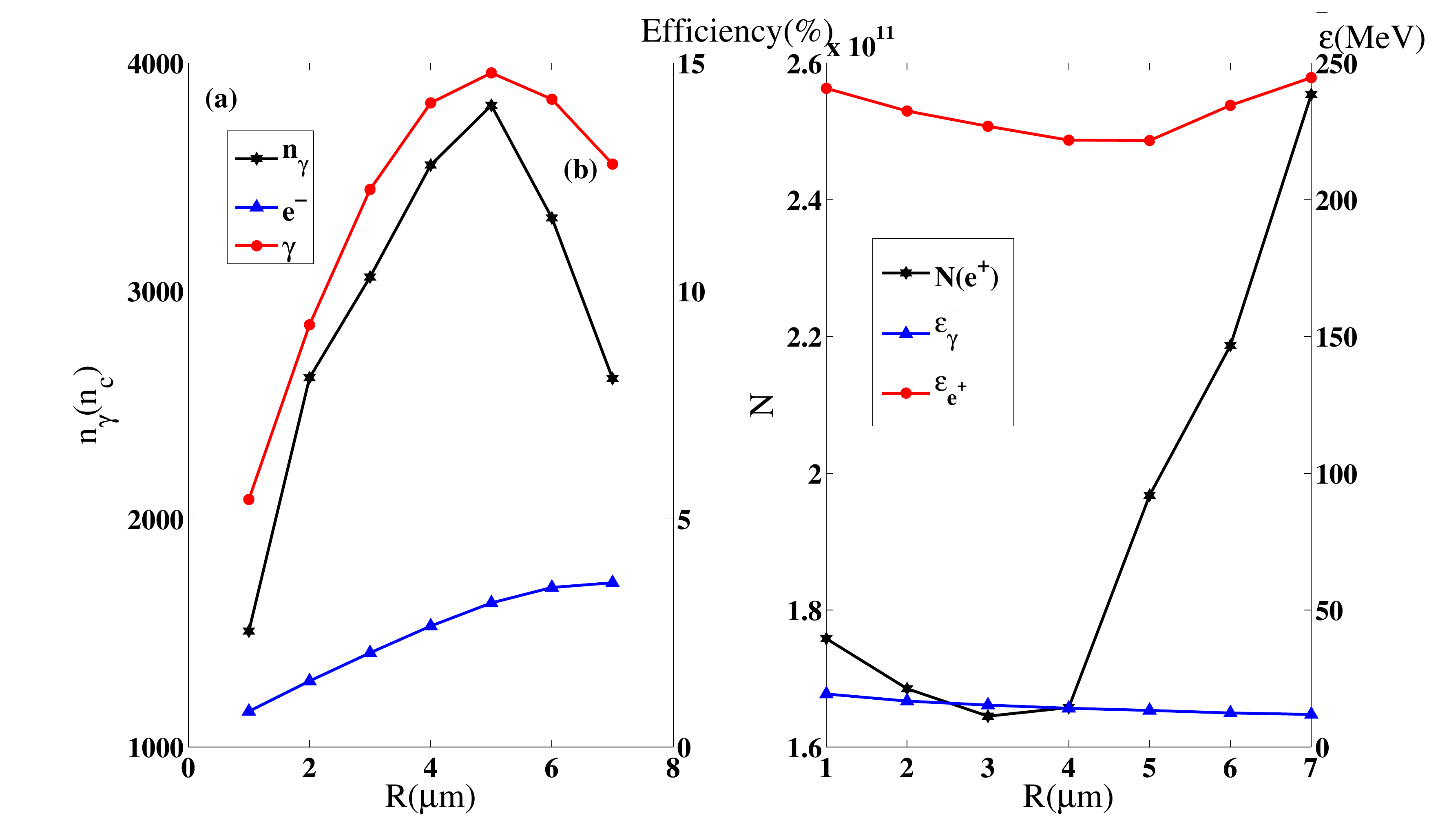}
\caption{\label{radius}(Color online) The peak number density of photons and laser energy efficiency to electrons and photons at $17 T_0$ with different R (a). The number of positrons, average energy of photons and positrons at $17 T_0$ with different R (b).}
\label{fig:7}
\end{figure*}

Figure \ref{radius}(a) shows the peak number density of $\gamma$-photons, laser-to-electrons and laser-to-photons energy conversion efficiency at $17 T_0$ with different target radius. On the one hand, the number density of photons is the highest when the radius is $5 ~\mathrm{\mu m}$, which is comparable to laser spot size. When the radius is small, the number and accelerating distance of electrons under target is reduced accordingly, which means a low energy of produced electrons resulting in a low rate for the  $\gamma$-ray emission. However, if the radius is large, the transverse Rayleigh-Taylor-like instability develops quickly \cite{Chen1,Wang1}, which will also lower the energy of electrons resulting in a undesirable $\gamma$-ray emission. So, the ultrabright $\gamma$-photons can be achieved when the laser field and circular structure collimate the electrons together. On the other hand, the laser-to-photons energy conversion efficiency is considerable when $R = 5 ~\mathrm{\mu m}$. It means the optimal target radius is $5 ~\mathrm{\mu m}$ for $\gamma$-ray emission when both the number and average energy of $\gamma$-photons are taken into account.

However, the average energy of $\gamma$-photons has a slow decrease as the target radius increases appropriately, as shown in Fig.\ref{radius}(b). The reason is that more high energy $\gamma$-photons are used for pair production by multi-photon BW process, although the electrons accelerating and high energy $\gamma$-ray emission become more significant as radius increases. As shown in Fig.\ref{radius}(b), when $R \ge 4 ~\mathrm{\mu m}$, the average energy of $\gamma$-photons decreases as $R$ increases. However, the number and average energy of positrons have a significant increase as $R$ increases when $R \ge 4 ~\mathrm{\mu m}$. It is obvious that the positron yield and average energy are almost the minimum value when the number density of photons and laser-to-photons energy conversion efficiency are maximum, comparing the Fig.\ref{radius}(b) with Fig.\ref{radius}(a). So, the circular target radius should be increased appropriately if it is designed for pair productions.

\subsection{The effect of incident laser beam mismatching}

In experiments, the deviation of incident lasers becomes a key issue for pair production. In order to check the influence of the mismatching of lasers on the $\gamma$-ray emission and pair production, we assume one laser is incident with a deviation $C$, where $C$ is a transverse offset compared to initial ideal case.

\begin{figure*}[htbp]\suppressfloats
\includegraphics[width=17cm, height=10cm]{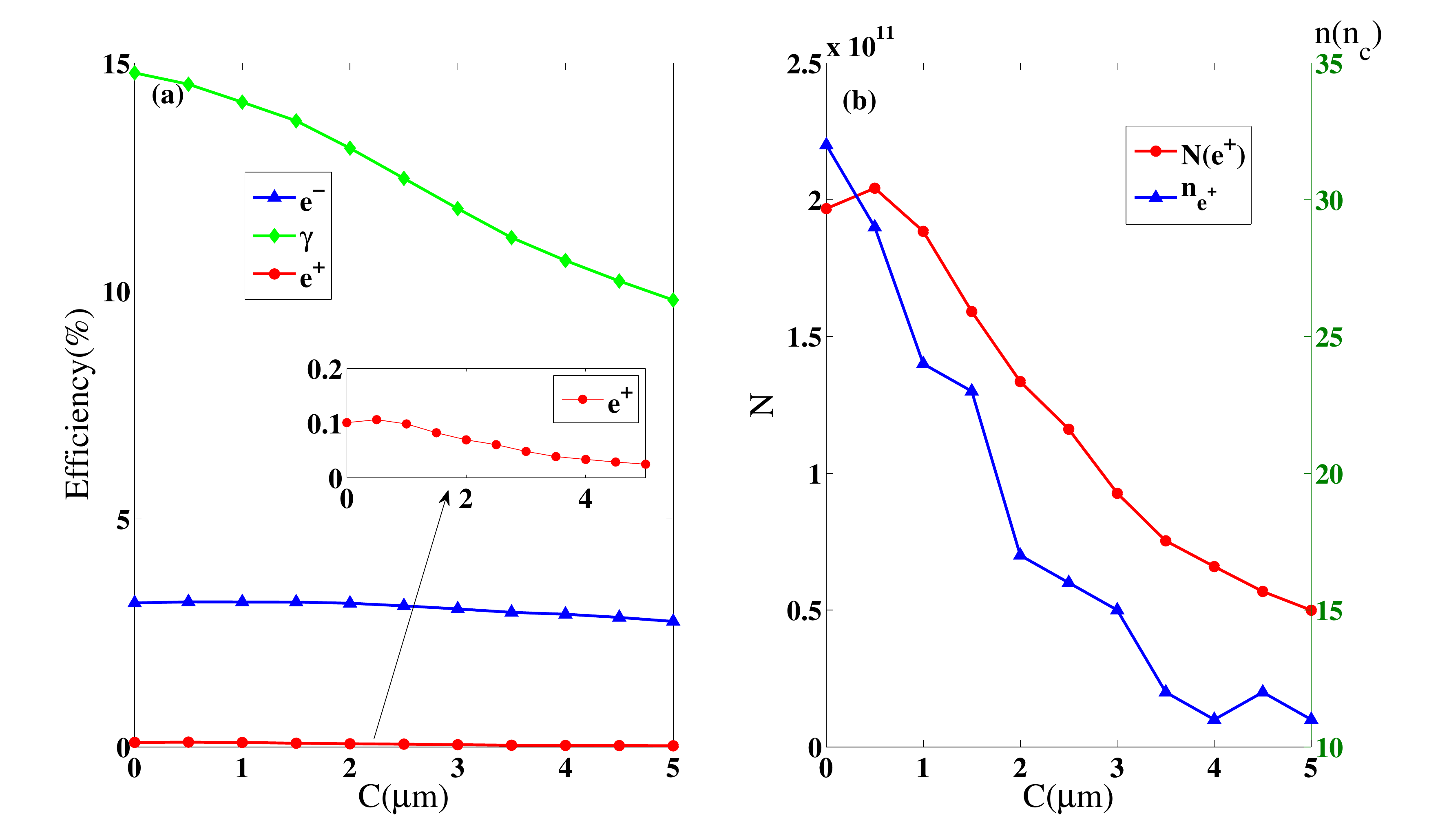}
\caption{\label{deviation}(Color online) The laser energy conversion efficiency to electrons, $\gamma$-photons and positrons at $17 T_0$ with different deviation C (a). The peak number density and number of positrons at $17 T_0$ with different deviation C (b). Here the number density is normalized by $n_c$.}
\label{fig:8}
\end{figure*}

The deviation of lasers will reduce the collision interaction of lasers and high energy $\gamma$-photons, which maybe decrease the probability of pair production. As Fig.\ref{deviation}(a) shows, the laser energy conversion efficiency to electrons, $\gamma$-photons and positrons at $17 T_0$ are diminished as $C$ increases. Besides, the number and peak density of positrons are also reduced as $C$ increases, as shown in Fig.\ref{deviation}(b). Above all, both the ultrabright $\gamma$-ray source and high quality positrons can be obtained if the deviation of lasers is controlled within $1 ~\mathrm{\mu m}$.

\section{Summary and conclusion}

In summary, a DLC circular target is proposed to replace the plane target and to enhance the $\gamma$-ray emission and $e^+e^-$ pairs production in present study by using the 2D3V QED-PIC code EPOCH. When two counter-propagating lasers are incident from the center of left and right boundary of the simulation box interact with the target, the circular target can enhance the laser-to-photons energy conversion efficiency by comparing the plane target. The density of $\gamma$-photons is increased about $2$ times of the plane target at $20 T_0$. Moreover, when another two counter-propagating lasers are incident from the center of up and down boundary of the simulation box, the overlap of multi-lasers will enhance the laser intensity and form a stable lattice-like optical trap. This optical trap can prevent the high energy electrons accelerated by RPA escaping from central interaction zone. Eventually, $7.5 \times 10^{14}$ $\gamma$-photons with average energy $16 ~\mathrm{MeV}$ is obtained and through NCBS, which is an order of magnitude higher than the photons yield from the plane target. The maximum density of $\gamma$-photons can be $5164 n_c$ at $14 T_0$, which may be ultrabright $\gamma$-ray source in the future application.

Compared with the two laser-driven circular target, we found the number and density of $\gamma$-photons has a nonlinear growth when another lasers is incident. These high quality photons collide with lasers resulting in above $20 n_c$ dense positrons via multi-photon BW process. As time goes on, the total positrons with average energy $230 ~\mathrm{MeV}$ yield can be $2.7 \times 10^{11}$. Furthermore, the optimal radius of circular target for $\gamma$-ray emission and pair production has also been analyzed and discussed respectively. For $\gamma$-ray emission, the optimal radius of target should be $5 ~\mathrm{\mu m}$. However, the radius should be increased suitably if one need more positrons. Lastly, the deviation of lasers is considered for real application, we found there is almost no effect on $\gamma$-ray emission and pair production if the deviation of lasers is controlled within $1 ~\mathrm{\mu m}$.

\section*{Acknowledgements}
This work was supported by the National Natural Science Foundation of China (NSFC) under Grant No. 11875007, and No. 11305010. Guoxing Xia's work is supported by the STFC Cockcroft Institute core grant. The computation was carried out at the HSCC of the Beijing Normal University. The authors are particularly grateful to CFSA at University of Warwick for allowing us to use the EPOCH.

\end{document}